# Non-local screening channels and metal-insulator transition in VO$_2$

R.J.O. Mossanek and M. Abbate*

*Departamento de Física, Universidade Federal do Paraná,*

*Caixa Postal 19091, 81531-990 Curitiba PR, Brazil*

We studied the changes in the electronic structure of VO$_2$ across the metal-insulator transition. The main technique was cluster model calculations with non-local screening channels. The calculation included a screening from a delocalized state at the Fermi level in the metallic phase, and a screening from the neighboring V site within the V-V dimmer in the insulating phase. Both the *coherent* and *incoherent* features in the V 3d band of metallic VO$_2$ are *well screened* states, whereas the true *poorly screened* 3d$^0$ state appears deeper in energy due to the relatively large U. The non-local screening state in the insulating phase involves U and consequently appears at higher energies. The change in the non-local screening channel opens the band gap originating the insulating phase.

Vanadium dioxide undergoes a first order metal-insulator transition [MIT] at about 340 K [1,2]. Above the transition temperature, $VO_2$ is a paramagnetic metal with a tetragonal symmetry. Below the critical temperature, it becomes a diamagnetic semiconductor with a monoclinic symmetry. The low temperature phase is characterized by the formation of distinct V-V dimers along the crystal c-axis. The electronic structure of $VO_2$ was first described in terms of molecular orbital theory [3]. This description was based on a single $V^{4+}$ ion surrounded by an edge sharing $O^{2-}$ octahedra. The crystal field splits the V 3d levels into three low energy $t_{2g}$ levels and two high energy $e_g$ levels. The lowest energy partially filled $t_{2g}$ level, which connects the V-V dimers, forms the so called $d_\parallel$ band. The V-V dimmerization splits the $d_\parallel$ band, forms a localized singlet, and opens a band gap.

$VO_2$ was studied using photoemission (PES) [4,5] and X-ray absorption (XAS) [6,7] spectroscopy. The electronic structure was studied using band structure [8-10] and cluster model [11,12] calculations. These early calculations were in qualitative agreement with the experimental results, and corroborated the main ideas of the molecular orbital model. But they failed to reproduce the value of the band gap, and the *coherent* and *incoherent* structures in the PES spectra [13]. A recent photoemission work [14], showed drastic changes in the *coherent* and *incoherent* structures across the MIT. In an effort to reproduced these features, many works were carried out using the GW [15] and DMFT [16,17] methods. These provided a better estimate of the band gap and reproduced the *coherent* and *incoherent* features. But they failed to explain key features in the unoccupied electronic states mapped by X-ray absorption [18].

We studied here the changes in the electronic structure of $VO_2$ across the MIT using cluster model calculations. Aside from the usual local ligand screening, we added a different non-local screening [19] channel to each phase. In the metallic phase, we included a screening from a coherent (long wavelength) Fermi level fluctuation [20]. We found that both *coherent* and *incoherent* structures present in the V 3d band are *well screened* states, while the *poorly screened* state appears at much higher energies. In the insulating phase, we added a screening from the V neighbor to account for intra-dimmer charge fluctuations. As a result, the V 3d occupied band is now formed by one *well screened* state, while the other is pushed close to the *poorly screened* state. In addition, this model reproduces the splitting observed experimentally in the $d_\parallel$ unoccupied band. This shows the close relationship between the screening channels and the metal-insulator transition.

The band structure was calculated using the Full Potential Linear Muffin Tin Orbital method [21]. The exchange and correlation potential was calculated using the Vosko approximation. The metallic phase was calculated in the tetragonal structure (space group P4$_2$/mnm). The lattice parameters were a = 4.530 Å and c = 2.896 Å and the atomic positions were V = (0.000, 0.000, 0.000) and O = (0.305, 0.305, 0.000). The self-consistent potential and the density of states were calculated using 24 irreducible **k**-points. The insulator phase was calculated in the monoclinic structure (space group P2$_1$/c). The lattice parameters were a = 5.743 Å, b = 4.517 Å, c = 5.375 Å and β = 121.56°. The atomic positions were V = (0.233, 0.024, 0.021), O1 = (-0.118, 0.288, 0.272) and O2 = (0.399, 0.315, 0.293). The self-consistent potential and the density of states were calculated using 80 irreducible **k**-points.

The cluster considered here consisted of a $V^{4+}$ ion surrounded by a regular $O^{2-}$ octahedra. The cluster model was solved using the standard configuration interaction method. The ground state was expanded in the $3d^1$, $3d^2\underline{L}$, $3d^3\underline{L}^2$, $3d^4\underline{L}^3$, $3d^5\underline{L}^4$ configurations (were $\underline{L}$ denotes an O 2*p* hole) [11,12]. The main parameters of the model were the charge-transfer energy Δ, the d-d Coulomb energy U, and the p-d transfer integral $T_\sigma$. The multiplet splitting was given in terms of the crystal field splitting 10Dq, the p-p transfer integral ppπ-ppσ, and the intra-atomic exchange J. The addition (removal) state was obtained by adding (removing) an electron to the ground state. Finally, the spectral weight was calculated using the sudden approximation. The main cluster model parameters were $\Delta = 2.0$ eV, $U = 4.5$ eV and $T_\sigma = 3.2$ eV. The multiplet splitting parameters were $J = 0.4$ eV, $10Dq = 1.9$ eV and $pp\pi-pp\sigma = 1.0$ eV. These parameters give the best results and are also in agreement with previous work [11,12].

The metallic phase calculation included a screening from a coherent state at the Fermi level. This multi-site screening was recently proposed in a hard X-ray photoelectron study of $V_2O_3$ [20]. The charge-transfer energy from the coherent state was $\Delta^*$ and the transfer integral was $T^*$ [20]. The metallic ground state was thus expanded in the $3d^1$, $3d^2\underline{C}$, $3d^2\underline{L}$, $3d^3\underline{C}^2$, $3d^3\underline{CL}$, and $3d^3\underline{L}^2$ configurations (were $\underline{C}$ denotes a hole in the coherent band) [20]. The charge-transfer energy from the coherent state in the metallic phase was set to $\Delta^* = 0.6$ eV. This energy is smaller than U because the charge comes from a long wavelength coherent state. The transfer integral involves V 3d-V 3d fluctuations at the Fermi level and was set to $T^* = 0.3$ eV. This effective value is in line with the V $t_{2g}$ bandwidth obtained from band structure calculations.

The insulating phase calculation included a screening from the neighboring $d_\parallel$ electron. This effect takes into account a single-site charge fluctuation within the V-V dimmer. This charge fluctuation costs U in energy and has a transfer integral of T'. The insulating ground state was then expanded in the $3d^1$, $3d^2\underline{D}$, $3d^2\underline{L}$, $3d^3\underline{DL}$, and $3d^3\underline{L}^2$ configurations (were $\underline{D}$ denotes a hole in the neighboring V site within the V-V dimmer). The charge fluctuation energy from the neighboring V site in the insulating phase was set to U. This value arises because, in this case, the screening charge comes from a single localized site. The transfer integral involves V 3d-V 3d fluctuations within the V-V dimmer and was set to T' = 2.5 eV. The larger value of T' reflects the larger overlap between the $d_\parallel$ states within the V-V dimmer.

Figure 1 shows the removal states of $VO_2$ decomposed in the different final state configurations. In the metallic phase, the *coherent* part, around –0.4 eV, is mostly formed by the $3d^1\underline{C}$ state, with 23 %, while the *incoherent* part, about –1.6 eV, is mainly formed by the $3d^1\underline{L}$ state, with 35 %. Both the *coherent* and *incoherent* parts correspond to the so called *well screened* states. The screening charge in the *coherent* part comes from a coherent state at the Fermi level, whereas the screening charge in the *incoherent* part comes from O 2p states in the ligand band. The so called *poorly screened* state, around –7.1 eV, is mainly composed by the $3d^0$ state, with 30%. The *well screened* states $3d^1\underline{C}$ and $3d^1\underline{L}$ appear close to the Fermi level because both $\Delta$ and $\Delta^*$ are relatively small, while the *poorly screened* state $3d^0$ appears deeper in energy because of the relatively large value of U.

In the insulating phase, the so called lower Hubbard band around –0.9 eV is actually formed by the $3d^1\underline{L}$ state with 29 %. This corresponds to a *well screened* state with the screening charge coming from O 2p states in the ligand band. The other *well screened* state in this case, $3d^1\underline{D}$, appears away from the Fermi level, around –5.5 eV. The screening charge of this state comes from the adjacent $d_\parallel$ electron within the V-V dimmer. The charge fluctuation from the single V neighbor costs U and pushes the state towards higher energies. Finally, the *poorly screened* state, which appears about –6.8 eV, is mostly composed by the $3d^0$ state. This state is the true lower Hubbard band and appears at higher energies because of the value of U.

These results show that $VO_2$ is in a heavily mixed charge-transfer regime, which was to be expected because $\Delta < U$ and T is relatively large. The *well screened* $3d^1\underline{L}$ states appear close to the Fermi level, around $\Delta$, whereas the *poorly screened* $3d^0$ state appears at higher energy, about U. This interpretation is qualitatively supported by a valence band resonant photoemission [22]. The differences between the metallic and insulating phase are mostly due to changes in the non-local screening channels. In the metallic phase, the screening from coherent states $3d^1\underline{C}$ appears close to the Fermi level, because the charge fluctuation $\Delta^*$ costs relatively little energy. In the insulating case, the screening from the neighboring V site $3d^1\underline{D}$ appears at higher energy, because the charge fluctuation costs a relatively large U. Finally, the usual screening from O 2p states $3d^1\underline{L}$ form a spectator state in the metallic phase, and becomes the lowest energy removal state in the insulating phase.

Figure 2 compares the results from the band structure and the cluster model calculations. As expected, the total DOS in the metallic phase is continuous at the Fermi level. The DOS in this region are mainly formed by V 3d states, although there are also considerable O 2p states from the covalent mixing. The V 3d states are split by the crystal field generated by the oxygen octahedra. The structure from –0.5 to 2.0 eV is related to the V $t_{2g}$ states while the structure from 2.5 to 5 eV corresponds to the V $e_g$ states. The $d_\parallel$ band, in this case, is mostly spread across the Fermi level from –0.5 to 1.0 eV. Finally, the charge fluctuations predicted by the band structure calculation are of the d-d type.

The cluster model calculation is a combination of the removal and addition states. The removal state presents two structures at –0.4 and –1.6 eV, which correspond to the *well screened* $3d^1\underline{C}$ and $3d^1\underline{L}$ states, respectively. The *poorly screened* $3d^0$ transition, which corresponds to the lower Hubbard band, appears deeper in energy. The addition state consists of various structures corresponding to the different $3d^2$ final state configurations. The structures at 0.6 and 1.0 eV represent the addition of a majority and minority $t_{2g}$ electron. The smaller peak, with a partial overlap at 1.0 eV, corresponds to the addition of a $d_\parallel$ electron. Finally, the structures at 3.3 and 3.7 eV represent the addition of a majority and minority $e_g$ electron. The charge fluctuations involve transitions from a coherent $3d^1\underline{C}$ state to a $3d^2$ state (the precise nature of the charge fluctuations is related to the orbital population in each phase [23]).

The total DOS in the insulating phase shows a semi-metallic character, with a pseudo-gap at the Fermi level. The absence of a true band gap is attributed to electron correlation effects beyond the LDA approach. The peak around –0.2 eV is related to the occupied part of the $d_\|$ band, whereas the structure at 2.0 eV corresponds to the unoccupied part of the $d_\|$ band. It is worth noting that the covalent O 2p character mixed in the $d_\|$ band is particularly small. The overall splitting of the $d_\|$ band, which is mainly caused by the lattice distortion within LDA, is about 2.0 eV. There are only minor changes in the $t_{2g}$ band, and the $e_g$ band becomes slightly broader in this phase. The charge fluctuations across the Fermi level continue to be of the d-d type, as in the metallic phase.

The first removal peak in the insulating phase, at –0.2 eV, corresponds to the *well screened* $3d^1\underline{L}$ state. As explained above, both the $3d^0$ and $3d^1\underline{D}$ states appear roughly at U below the Fermi level. The intra-dimmer screening, through the T' hybridization, pushes down the $t_{2g}$ and $e_g$ addition states. But the $d_\|$ addition state is not affected because it corresponds already to a doubly occupied configuration. The combined result is a shift of the $d_{//}$ addition state to about 1.9 eV towards higher energies. The calculated band gap, approximately 0.9 eV, is in good agreement with the experimental results. The lowest energy charge fluctuations involve transitions from a $3d^1\underline{L}$ to a $3d^2$ state, and the band gap is consequently of the charge-transfer p-d type.

Figure 3 compares the photoemission (PES) spectra, from Ref. 4, to the cluster model calculation. The results of the calculation are in very good agreement with the experimental spectra. The energy region in the spectra exhibits the changes in the V 3d band across the transition (the O 2p band appears deeper in energy and does not change much during the transition). In the metallic phase, the spectrum presents the coherent screening peak $3d^1\underline{C}$ around –0.4 eV and the ligand screening part $3d^1\underline{L}$ about –1.6 eV. In the insulating phase, the ligand screening peak appears around –0.9 eV, and the intra-dimmer screening part deeper in energy. The absence of the coherent part close to the Fermi level in the insulating phase opens the band gap. This shows the relationship between the screening channels and the metal-insulator transition in $VO_2$.

Figure 4 compares the O 1s X-ray absorption (XAS) spectra, from Ref. 6, and the cluster model calculation. The calculated spectra are again in good agreement with the experimental data. In the metallic phase, the structure around 1.0 eV is related to the $t_{2g}$ band and the structure around 3.5 eV to the $e_g$ band (the larger intensity of the $e_g$ band is due to the stronger hybridization with the O 2p states). The $d_\parallel$ band, in the metallic phase, appears around 1.0 eV partially overlapping the $t_{2g}$ band. In the insulating phase, the $d_{//}$ band is shifted approximately to 1.9 eV, towards higher energies. On the other hand, the $t_{2g}$ band does not change and the $e_g$ band presents only a minor shift to higher energies. The calculation of the related inverse photoemission spectra exhibits a similar evolution of the spectral weight [24].

In conclusion, we studied the changes in the electronic structure of $VO_2$ across the metal-insulator transition. These changes are mainly attributed to differences in the non-local screening processes in each phase. The screening from a coherent state $3d^1\underline{C}$ appears next to the Fermi level (about $\Delta^* \approx 0.6$ eV) in the metallic phase; whereas the screening from the dimmer neighbor $3d^1\underline{D}$ in the insulating phase appears deeper in energy (around $U \approx 4.5$ eV). This transfer of spectral weight is responsible for the opening of the band gap in the insulating phase. The structures close to the Fermi level are *well screened* states, while the true *poorly screened* state appears at higher energies. These results confirm that the $VO_2$ compound is in a highly covalent (large T) charge transfer regime ($\Delta < U$).

**References**

*Corresponding author: miguel@fisica.ufpr.br

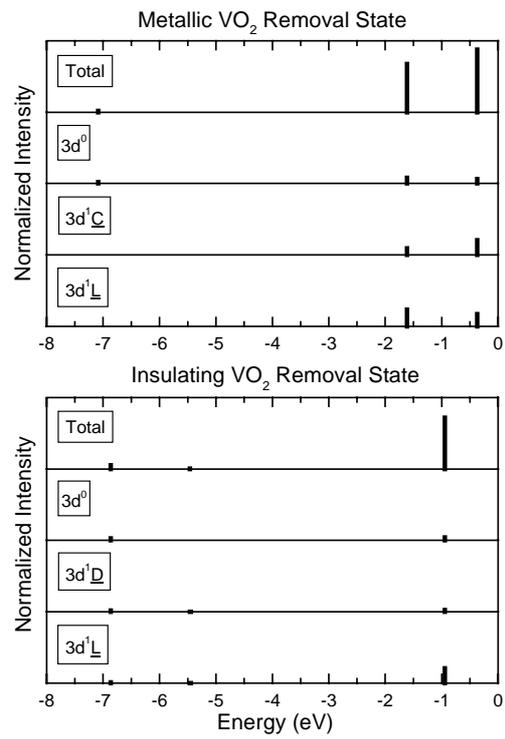

**Figure 1:** Removal states of metallic and insulating $VO_2$ decomposed on the main final state configurations.

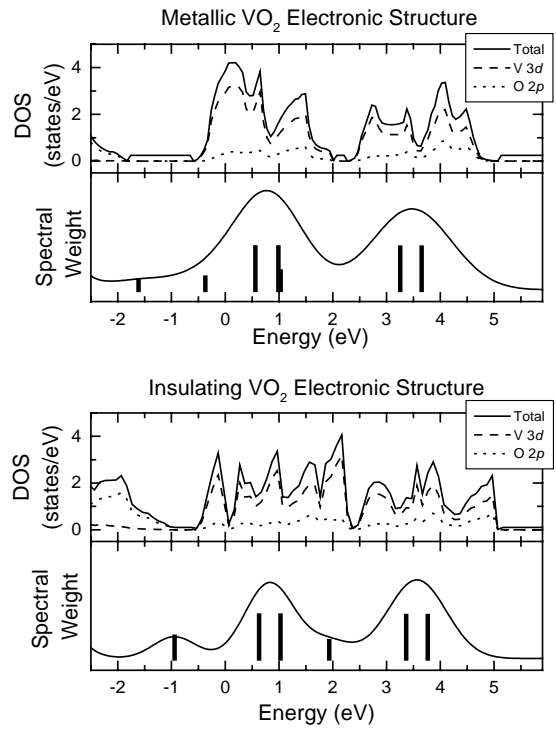

**Figure 2:** Cluster model calculations of metallic and insulating VO$_2$ compared to band structure calculations.

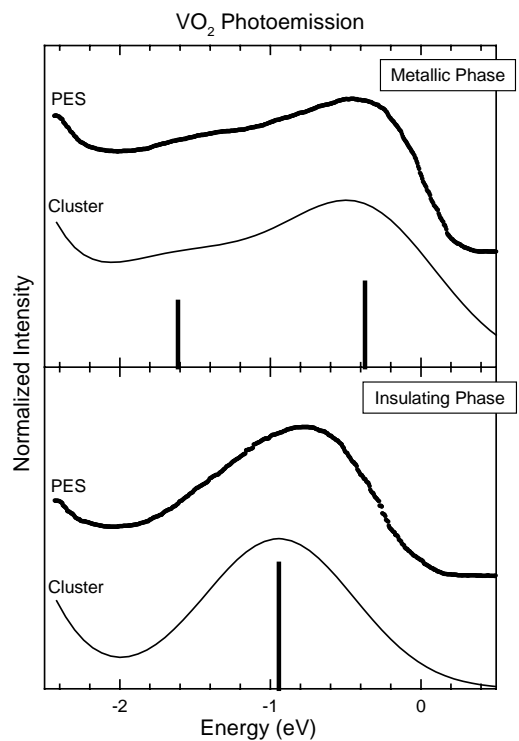

**Figure 3:** Cluster model calculations of metallic and insulating VO$_2$ compared to the photoemission (PES) spectra taken from Ref. 4.

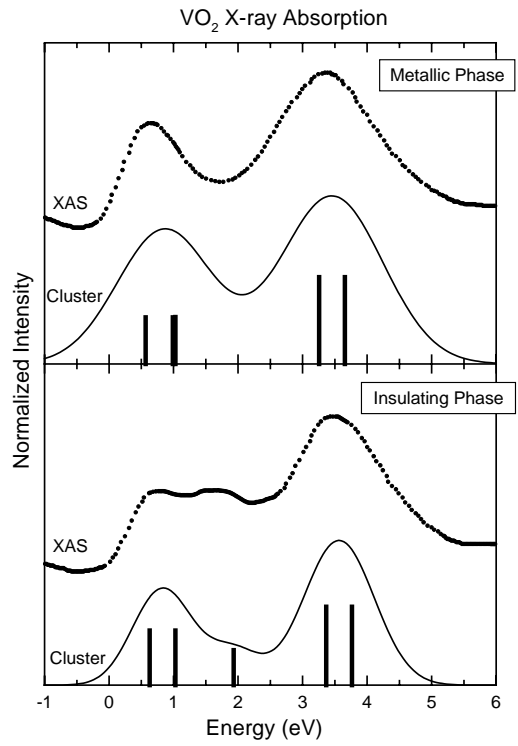

**Figure 4:** Cluster model calculations of metallic and insulating VO$_2$ compared to the X-ray absorption (XAS) spectra taken from Ref. 6.